\documentclass{article}

\PassOptionsToPackage{numbers}{natbib}

    \usepackage[preprint]{cs285}

\usepackage[utf8]{inputenc} 
\usepackage[T1]{fontenc}    
\usepackage{hyperref}       
\usepackage{url}            
\usepackage{booktabs}       
\usepackage{amsfonts}       
\usepackage{nicefrac}       
\usepackage{microtype}      

\usepackage{graphicx}
\usepackage{caption}
\usepackage{subcaption}
\usepackage{float}
\usepackage{xcolor}
\usepackage{amsmath}
\usepackage{enumitem}

\title{Safe Deep Model-Based Reinforcement Learning with Lyapunov Functions}

\author{%
  {\ Harry Zhang}\\
    Course Project for 16.332\\
}

\newcommand\mymathop[1]{\mathop{\operatorname{#1}}}
\begin{document}

\maketitle
\begingroup
\addtolength\leftmargini{-0.3in}
\textbf{Summary: }Reinforcement learning (RL) for control systems with unknown nonlinear dynamics and safety constraints has been difficult due to unintended behavior from imperfect cost function and dynamic uncertainty, which could lead to instability breaking down the system. We propose a new RL framework to enable efficient policy learning with unknown dynamics based on the recently proposed learning model predictive control framework \cite{rosolia2017learning,rosolia2017robust,rosolia2019sample, pan2022tax, zhang2016health}, with provable guarantees of stability. The new stability-augmented framework consists of a neural-network-based learner that learns a Lyapunov function, and a model-based RL agent to consistently complete the tasks while satisfying user-specified constraints given only sub-optimal demonstrations and sparse-cost feedback.

Our work is based on Safety Augmented Value Estimation from Demonstrations (SAVED) \cite{thananjeyan2020safety, sim2019personalization, elmquist2022art, zhang2020dex, zhang2021robots}. SAVED is a sample-efficient method that can consistently complete tasks with complex safety constraints using only non-expert, sub-optimal demonstrations and sparse rewards from simple supervision identifying task completion. By adding safety constraints during the training of policy learning, SAVED ensures a high probability of being able to plan back to the safe region. However, it has no stability guarantees as the learned value function is an arbitrary neural network that is not constrained to be a Lyapunov function. To address the stability issues which could cause states to blow up and catastrophic consequences in practice, we use the learned Lyapunov network for the model-based RL agent as the value function, which will act in the environment in tandem with the safety-augmented agent in order to achieve both stability and safety.

To construct and learn a Lyapunov-constrained value function approximator, we train a Bayesian Neural Network with a loss function adapted from \citet{mittal2020neural, zhang2023flowbot++, eisner2022flowbot3d, avigal20206, avigal2021avplug}. The output from the network is defined as $V_{Net}(x)$, a $n_V \times n_x$ matrix, and the actual Lyapunov function used as the value function can be derived as $V(x) = x^T(l_lI+V_{net}(x)^TV_{net}(x))x$, where $x$ is the state and $l_l$ is a tunable parameter.

We train a 3-layer neural network with 500 hidden units per layers and swish activations, as in \cite{thananjeyan2020safety, pan2023tax, lim2021planar, lim2022real2sim2real} and evaluate the performance of the SAVED algorithm with and without Lyapunov-constrained value functions in four simulated environments and point-mass navigation tasks. The effectiveness of our method and its effects on the stability during training as well as the optimality of the learned policy are shown and summarized in \autoref{fig:iteration-cost}. From the results, we found that the Lyapunov value function significantly improved the controller performance by training a value function that was monotonically decreasing along the planning trajectories, which prevented the agent from being stuck in local minima around the initial states and resulted in more optimal trajectories with less variance across runs as well as higher task completion and constraint satisfaction rates.

Our contribution with this study is twofold: first, the implementation of learning a Lyapunov function as the value function of SAVED; second, the empirical validation of the theoretical performance improvement in terms of stability, task completion rates, and constraint satisfaction rates. Some future work of this study includes the physical experiments on a robot and the study of stronger stability guarantees such as asymptotic stability.
\endgroup

\clearpage
\begin{abstract}
Model-based Reinforcement Learning (MBRL) has shown many desirable properties for intelligent control tasks. However, satisfying safety and stability constraints during training and rollout remains an open question. We propose a new Model-based RL framework to enable efficient policy learning with unknown dynamics based on learning model predictive control (LMPC) framework with mathematically provable guarantees of stability. We introduce and explore a novel method for adding safety constraints for model-based RL during training and policy learning. The new stability-augmented framework consists of a neural-network-based learner that learns to construct a Lyapunov function, and a model-based RL agent to consistently complete the tasks while satisfying user-specified constraints given only sub-optimal demonstrations and sparse-cost feedback. We demonstrate the capability of the proposed framework through simulated experiments. Code and data are available \href{https://github.com/harryzhangOG/salved}{\textcolor{blue}{here}}.

\end{abstract}

\section{Introduction}
Modern control problems require a proper understanding of the system. However, the dynamic model of some tasks could be very complicated with thousands of parameters which makes it impossible to deploy system identification. Furthermore, some tasks require additional safety constraints such as the collision avoidance in navigation and manipulation, velocity and position limitation in human robot interaction, which makes the problem harder. In addition, a poorly designed controller could lead to unstable behavior resulting in catastrophic consequences.

Deep model-based reinforcement learning (MBRL) has the advantage of using a deep neural network to express the dynamic model with sample efficiency. However, the performance of deep model-based RL usually depends on the design of the cost function which could lead to unintended behavior \cite{thananjeyan2020safety, devgon2020orienting, yao2023apla, shen2024diffclip}. Prior work has been done to relax the requirement of the cost function design by using the task completion as the supervision with some sub-optimal demonstrations~\cite{thananjeyan2020abc,thananjeyan2020safety}. To ensure the agent staying in the safety region, safety constraints are added during the training the RL agent.

One concern of deep model-based RL is the stability of the closed-loop system with the learned policy. Since the dynamic model is not perfectly modeled and the outputs from the learned policy are not explicitly constrained, there is no stability guarantee. A common verification of stability of the dynamic system is the existence of the Lyapunov function, which implies the convergence to the equilibrium point of the resulting trajectory. A Lyapunov function for a system with unknown dynamics can be estimated using a neural network \cite{berkenkamp2017safe}. \cite{bobiti2016samplingbased,bobiti2017,gallieri2019safe} have shown that the learned Lyapunov neural network can provide stability guarantees and improve the performance.

In this study, we focus on policy learning for unknown systems with safety constraints and stability guarantees. We use a learned Lyapunov function as the value function for Safety Augmented Value Estimation from Demonstrations (SAVED)~\cite{thananjeyan2020safety}, a learning method which is able to complete the tasks efficiently with safety constraints using simple task completion supervision and sub-optimal demonstrations. The Lyapunov function is learned using a neural network with a loss function proposed in \cite{mittal2020neural}. Our experiments using the Safety Augmented Lyapunov Value Estimation from Demonstrations (SALVED) show the performance improvement brought by the Lyapunov value function, which helps to increase the stability of the MPC controller and lower the probability of the controller getting stuck in sub-optimal trajectories by forcing the trained value function to be monotonically decreasing along the closed-loop trajectories and reducing the number of local minima in the value function. The results from SALVED also show higher task completion and constraint satisfaction rates and more optimal trajectories with less variance.

The contributions of this study consist of two parts: (1) the implementation of the Lyapunov value function in SAVED and (2) the experiments validating the benefits and stability of the new framework SALVED.
\section{Related Work}
Significant interest in model-based planning and
data-driven model-based RL \cite{chua2018deep, deisenroth2011pilco, fu2016one, lowrey2018plan, nagabandi2018neural} has been sparked in research due to the improvements in sample efficiency when planning over learned dynamics compared to model-free methods for continuous control \cite{fujimoto2018addressing, haarnoja2018soft}. Moreover, model predictive control (MPC) with learned model through data-driven approach has been shown to be promising as it has been applied in various tasks such as deformable objects manipulation \cite{yan2020learning}, and car racing \cite{rosolia2017learning}.

The safety of a control system implies the ability of the system to stay in the safe region and converge to the region of attraction. In reinforcement learning, the safety of the closed loop system with agent is a challenging problem due to unknown system dynamics. Safe RL has risen broad interests in the past few decades \cite{garcia2015comprehensive}. \cite{thananjeyan2020abc,thananjeyan2020safety} focuses on the problem of ensuring the states lie in safety set with high probabilities for system with unknown dynamics. \cite{thananjeyan2020abc,thananjeyan2020safety} solve the problem using sparse cost and probabilistic constraints during exploration. \cite{achiam2017constrained,li2018safe} study the problem of exploration with explicit constraints satisfied.

The study on the stability of closed loop system with RL agents mostly adapts the idea of using Lyapunov stability criterion. To our knowledge, \cite{perkins2002Lyapunov} first used Lyapunov function in reinforcement learning to guarantee the closed loop system stability. The construction of the Lyapunov function for an unknown system is challenging. \cite{berkenkamp2017safe} used a neural network to estimate the Lyapunov function. \cite{chow2018Lyapunov} proposed a Lyapunov based approach for constrained Markov decision process (CMDP) by constructing the Lyapunov function using an LP-based algorithm. \cite{mittal2020neural} proposed a method using a learned Lyapunov Neural Network as the value function estimation to be the terminal cost of the Model predictive control (MPC) and showed an increased stable horizon compared to the baseline MPC. \cite{gallieri2019safe} learned the safety set from the Lyapunov function. The Lyapunov network proposed in \cite{gallieri2019safe} is used in \cite{mittal2020neural}. In this paper, we use the same Lyapunov network from \cite{gallieri2019safe} to learn a Lyapunov function as the value function.

\section{Preliminaries}
In this section, we introduce some key concepts that this paper builds upon.
\subsection{Control Lyapunov Functions}
A Lyapunov function $V(x)$ is a positive scalar function that takes as input a state and outputs a positive scalar value:
\[
V(x):\; \mathbb{R}^n_x\to\mathbb{R}^+
\],
and it should satisfy the following constraint that it monotonically decreases along the closed-loop trajectory of a controlled system until a target point (or set) is reached. The existence of such a function is a necessary and sufficient condition for stability and convergence of a dynamical system~\cite{khalil2014nonlinear}. Control Lyapunov functions  can be used to design control policies.

A control Lyapunov function should be upper and lower bounded by strictly increasing, unbounded,
positive functions $\mathcal{K}_\infty$~\cite{khalil2014nonlinear}. Consider the optimal control problem with cost-to-go function defined as:
\[
c(x, u) = x(t)^TQx(t) + u(t)^TRu(t)
\],
where $x$ is the state and $u$ is the input. The $\mathcal{K}_\infty$ can be designed as the scaled, squared $l_2$ norm of the state:
\[
l_l||x(t)||_2^2\leq V(x(t))\leq L_V||x(t)||_2^2
\]
where $l_l$ is the minimum eigenvalue of $Q$ and $L_V$ is a local Lipschitz constant for $V$.

Furthermore, we want the Lyapunov function values to decrease monotonically along the closed-loop system's trajectory. The amount of the decrease is usually chosen to be quantified by the stage cost (cost-to-go) value incurred at the current time step. Mathematically, this condition can be written as:
\[
V (x(t + 1)) - V (x(t)) \leq -c(x(t), u(t)).
\]

\subsection{Model Predictive Control}
Model predictive control (MPC) is a control strategy that iteratively plans ahead and solves for a sequence of actions to take that incurs the minimal cost-to-go values in the plan horizon. The agent will only take the first action in the planned sequence in order to minimize the deviation caused by model inaccuracy. The iterative optimization problem that MPC tries to solve is described as follows:
\begin{align*}
    J^*(x(t)) &= \min_{u_t\dots u_{t+N-1}} \sum_{k=0}^{N-1}c(x_{t+k}, u_{t+k})\\
    \text{s.t. }x_{t+k+1} &= f(x_{t+k}, u_{t+k})\\
\end{align*},
where $N$ is the planning horizon of the MPC and we only take the first element in the planned the sequence of inputs $u_t\dots u_{t+N-1}$. After taking this step, we result in the next state and re-plan according to the above optimization problem.

\section{Methods}
\begin{figure}
\begin{subfigure}{.24\textwidth}
\includegraphics[width=\linewidth]{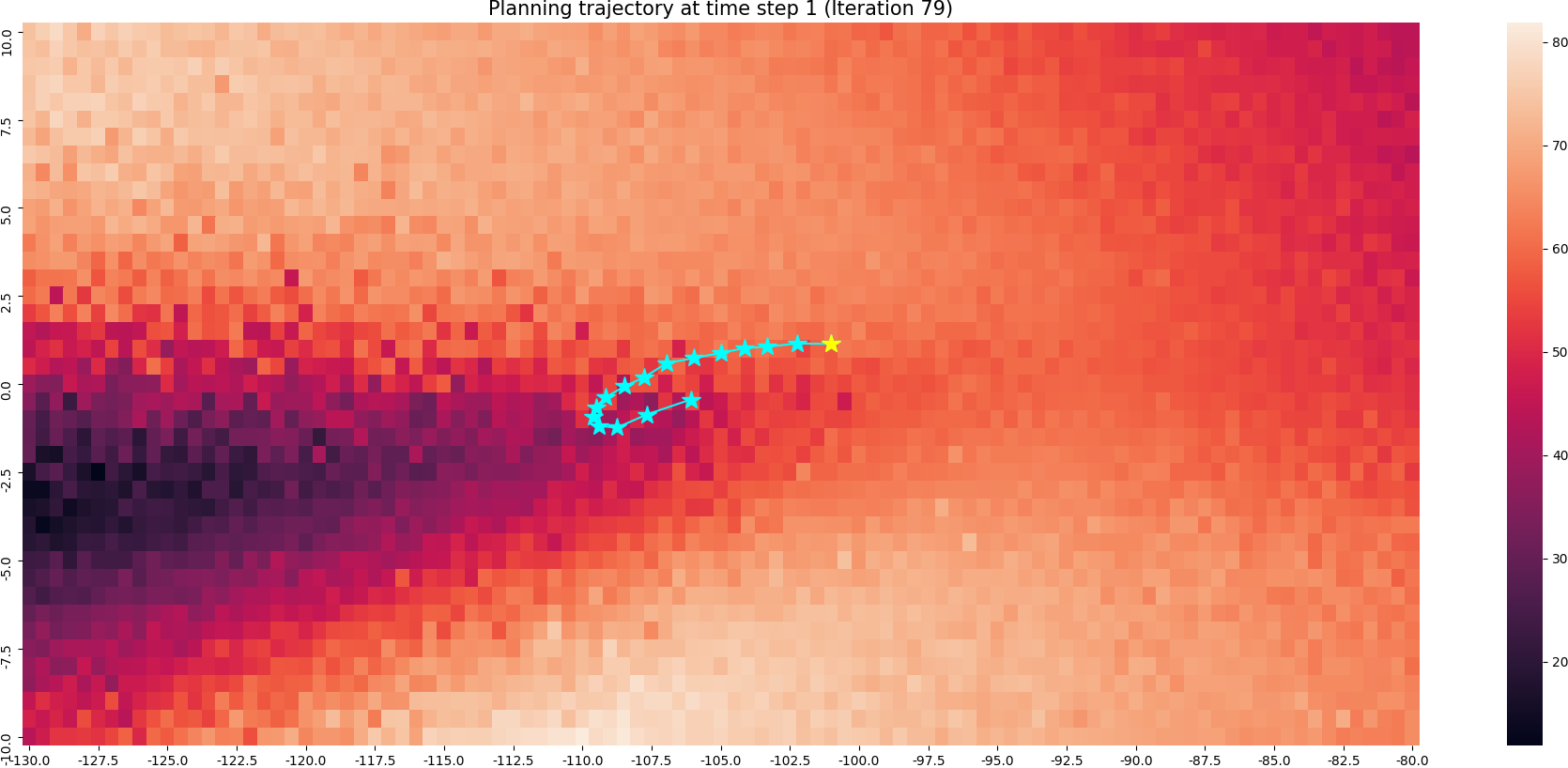}
\end{subfigure}\hfil
\begin{subfigure}{.24\textwidth}
\includegraphics[width=\linewidth]{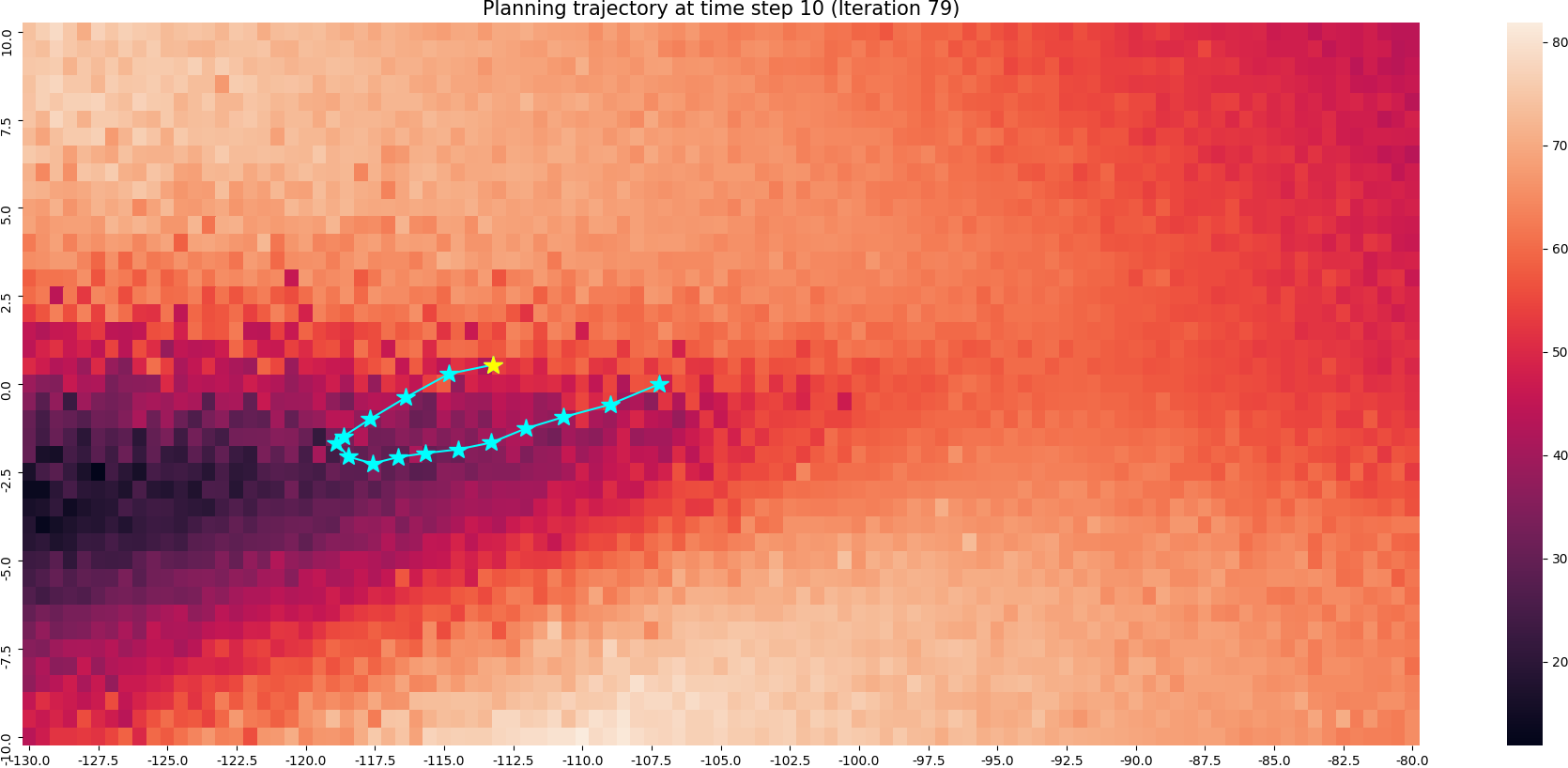}
\end{subfigure}\hfil
\begin{subfigure}{.24\textwidth}
\includegraphics[width=\linewidth]{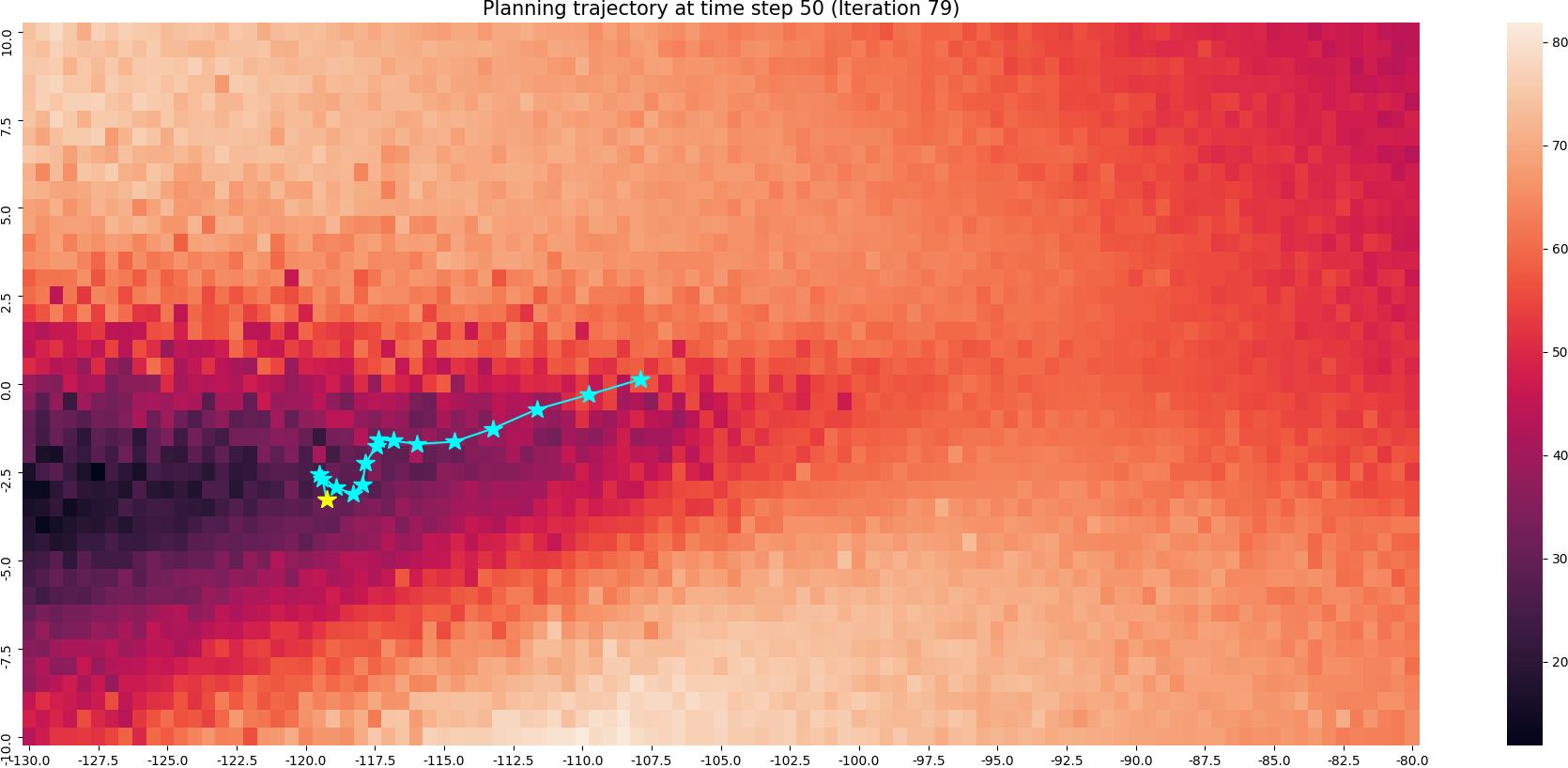}
\end{subfigure}\hfil
\begin{subfigure}{.24\textwidth}
\includegraphics[width=\linewidth]{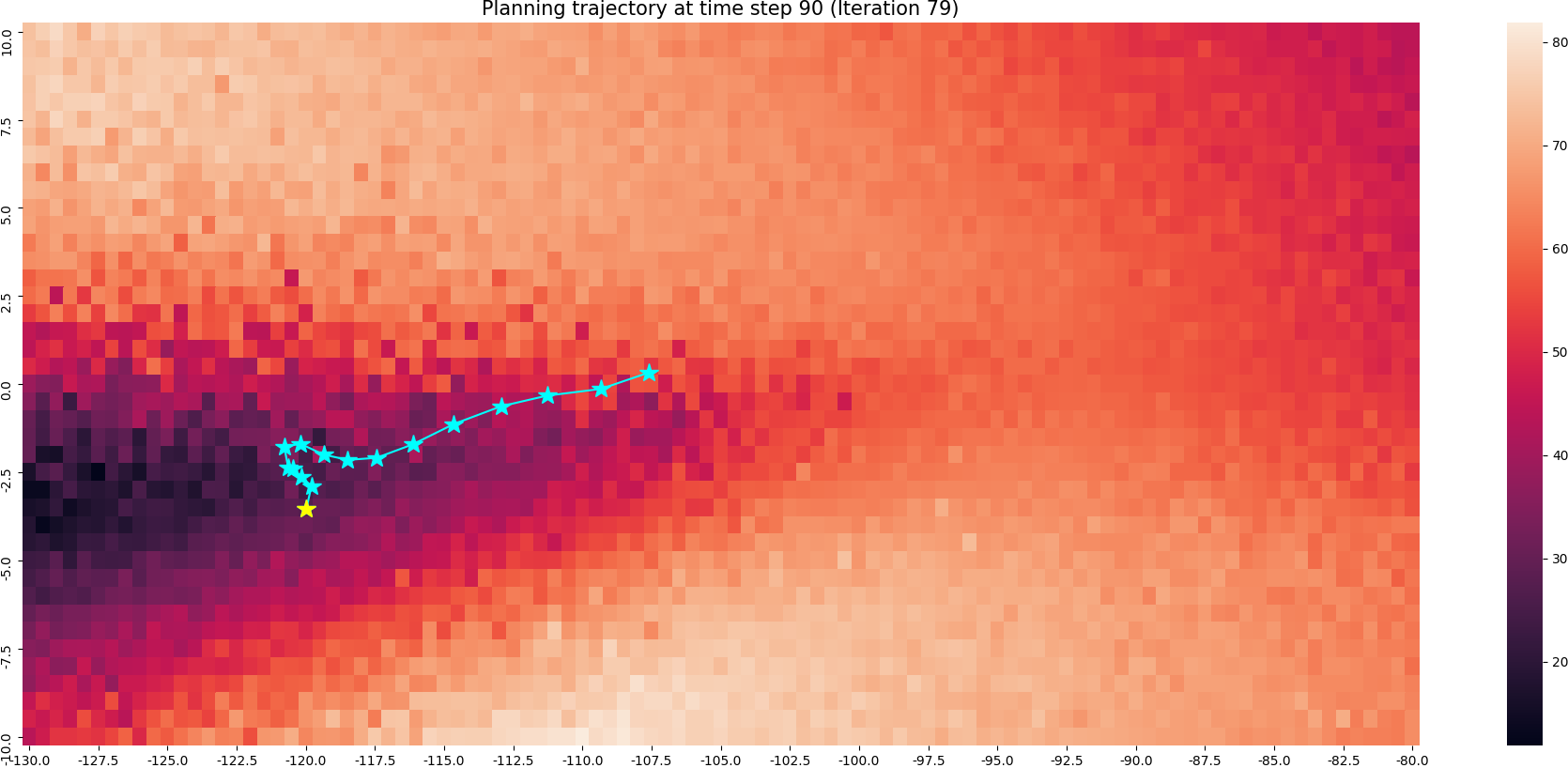}
\end{subfigure}
\caption{Planning trajectories of the MPC policy in SAVED in Pointbot 1 (point mass navigation) environment, where the value of the heat map is the predicted cost-to-go (negative returns) according to the value function at each iteration. In the plots, the $x$-axis is [-130, -80] and the $y$-axis is [-10, 10]. The starting position $x_0$ is $(-100, 0)$ and the goal is at the origin, $(0, 0)$. We see that even in an environment with no obstacles or constraints, SAVED struggles to complete the task due to local minima.}
\label{fig:pb1-local-minima}
\end{figure}

The main novelty of our work is that we are trying to design a Lyapunov-constrained value function which the model-based RL agent can leverage in order to achieve stable performance. Specifically, we plan to improve the SAVED algorithm, which can perform poorly if the learned value function contains many local minima as shown in \autoref{fig:pb1-local-minima}, using a Lyapunov neural network that can guide the agent to correctly plan the action to take at each time step.

\subsection{SAVED Algorithm}
At each time step, SAVED solves the optimization problem in \autoref{eq:saved}, where $C(x_{t+i},u_{t+i})$ is the sparse cost function indicating task completion, $V_\phi^\pi(x_{i+H})$ is the value function and $f_\theta$ is the learned dynamic model. $\rho_\alpha$ is the safety density model constraining the exploration by requiring $x_{t+H}$ to be in the region with high task completion probability. And $\beta$ is the probability of $x_{t:t+H}$ falling in the feasible region $\mathcal{X}^{H+1}$, where $\mathcal{X}^{H+1}$ is the set of $H+1$ length state sequences \cite{berkenkamp2017safe}.
\begin{equation}
    \begin{aligned}
   & u^*_{t:t+H-1} =\displaystyle\mymathop {argmin}_{u_{t:t+H-1}\in\mathcal{U}^H}\mathbb{E}_{x_{t:t+H}}[\sum_{i=0}^{H-1}C(x_{t+i},u_{t+i})+V_\phi^\pi(x_{i+H})],\\
    &\text{s.t.}\quad x_{t+i+1}\sim f_\theta(x_{t+i,u+i}) \forall i \in \{0,\dots,H-1\},\\
     &\quad\quad\rho_\alpha(x_{t+H})>\delta,\mathbb{P}(x_{t:t+H}\in \mathcal{X}^{H+1})\geq \beta
    \label{eq:saved}
\end{aligned}
\end{equation}

\subsection{Lyapunov Value Function}
We plan to constrain the value function used by the model-based RL to be a control Lyapunov function. To achieve this, we use the function introduced in \cite{gallieri2019safe}:
\begin{equation}
s_v V (x) = s_v \cdot x^T\left(l_lI + V_\text{net}(x)^TV_\text{net}\right)x,
\label{eq:lyv}
\end{equation}

where $V_\text{net}$ is a neural network that
produces a $n_V \times n_x$ matrix. Here $n_V$, $s_v$, and $l_l>0$ are hyper-parameters. In the original introduction of the value function in \cite{gallieri2019safe}, $s_v$ is fixed to be 1, but in our experiments, we have found that with $s_v = 1$, $V(x)$ tends to be numerically unstable, causing the loss to blow up, so we scale down the value function output significantly (e.g. $s_v = 0.001$).

We use the same loss function as described in \cite{mittal2020neural}, which enforces the value function to be control Lyapunov. The objective of the agent is to minimize the following loss in order to train the Lyapunov neural network:
\[
\min_{V_\text{net}, l_s} \mathbb{E}_{(x, u, x')}\left[J(x, u, x')\right]
\]
where $J(x, u, x')$ is defined as
\[
J(x, u, x') =\frac{\mathcal{I}(x)}{\rho}J_s(x, u, x') + J_\text{vol}(x, u, x')
\]
and
\begin{align*}
    \mathcal{I}(x) &= 0.5(\text{sign}[l_s-V(x)] + 1)\\
    J_s(x, u, x') &= \frac{\text{ReLU}[\Delta V(x)]}{V(x) +\epsilon}\\
    J_\text{vol}(x, u, x') &=\text{sign}[\Delta V(x)](l_s - V(x))\\
    \Delta V(x) &= V(x') - \lambda V(x) + v\mathcal{I}(x)
\end{align*}.

In the definitions above, $\mathcal{I}(x)$ is an indicator function representing the step cost that the agent incurs at a given state $x$. $J_\text{vol}$ is a classification loss that computes the correct boundary between the stable and unstable points, which also helps to increase the safe set volume \cite{mittal2020neural}. The scalars $\epsilon > 0$, $\lambda\in[0, 1)$, $v\in[0, 1]$, and $\rho > 0$ are hyper-parameters. $l_s$ is a trainable variable, and the role of $l_s$ is to provide an upper bound for $V(s)$ such that the states can remain in an invariant set.

\subsection{Safety-Augmented Lyapunov Value Estimation from Demonstration (SALVED)}
By substituting the Lyapunov value function $V$ calculated from \autoref{eq:lyv} in \autoref{eq:saved}, we have SALVED to be SAVED with a Lyapunov value function, which provides an estimate for the terminal cost for the current policy at the end of the planning horizon:
\begin{equation}
    u^*_{t:t+H-1} =\displaystyle\mymathop {argmin}_{u_{t:t+H-1}\in\mathcal{U}^H}\mathbb{E}_{x_{t:t+H}}\left[\sum_{i=0}^{H-1}C(x_{t+i},u_{t+i})+s_v V(x_{i+H})\right],\\
\end{equation}
According to Lemma 1 from \cite{mittal2020neural}, for a given horizon length $N$ and contraction factor $\lambda$, there is minimum scaling of Lyapunov function $V$, which is calculated from \autoref{eq:lyv}, and a lower bound on the discount factor $s_v$ such that the system under MPC is stable.

When the learned dynamic model $f_\theta$ is perfect, the state would converge to the equilibrium point. In practice, however, the learned dynamics model may be imperfect for various reasons, such as due to aleatoric uncertainty in the actual dynamics and the epistemic uncertainty as a result of insufficient information about the environment, but the system can still stable when error between the learned dynamic model and the actual model is bounded, in which case the state would converge to a bounded area around the equilibrium point, whose size depends on the model error, $N$ and $s_v$ \cite{mittal2020neural}.

\section{Experiments}

\subsection{Experimental Setup}

For all of our SALVED experiments, we used Safety-Augmented Value Estimation from Demonstrations (SAVED) as our baseline. In order to test whether the new Lyapunov-based value functions can provide additional stability to model-based RL methods that are based on the learning MPC framework, such as SAVED, we benchmarked our proposed method, SALVED, and compared its performance with that of the baseline on four 4-dimensional simulated navigation tasks in which a point-mass navigates from a random starting position to a goal set, which is defined as a unit ball centered around the origin:

\begin{itemize}[leftmargin=*, itemsep=0pt, topsep=0pt]
  \item \textbf{Pointbot 1}: long navigation task with no obstacles
  \item \textbf{Pointbot 2}: medium-length navigation task with a single large obstacle along the $y$-axis
  \item \textbf{Pointbot 3}: medium-length navigation task with two medium-sized obstacles and an open channel along the $x$-axis
  \item \textbf{Pointbot 4}: long navigation task with obstacles of varying sizes surrounding the goal and a small opening for entry
\end{itemize}

\subsection{Implementation Details}

The dynamics models for the baseline and SALVE have the same network architecture, which is a probabilistic ensemble of five 3-layer neural networks with 500 hidden units per layer and swish activations. For SAVED, the value function has the same network architecture as its dynamics model, whereas for SALVED, the value function is a neural network with the same number of layers, same number of hidden units per layer, and the same activations but constrained to be a Lyapunov function.

\subsection{Results}

\begin{figure}[H]
  \centering
  \begin{subfigure}{0.5\textwidth}
  \includegraphics[width=\linewidth]{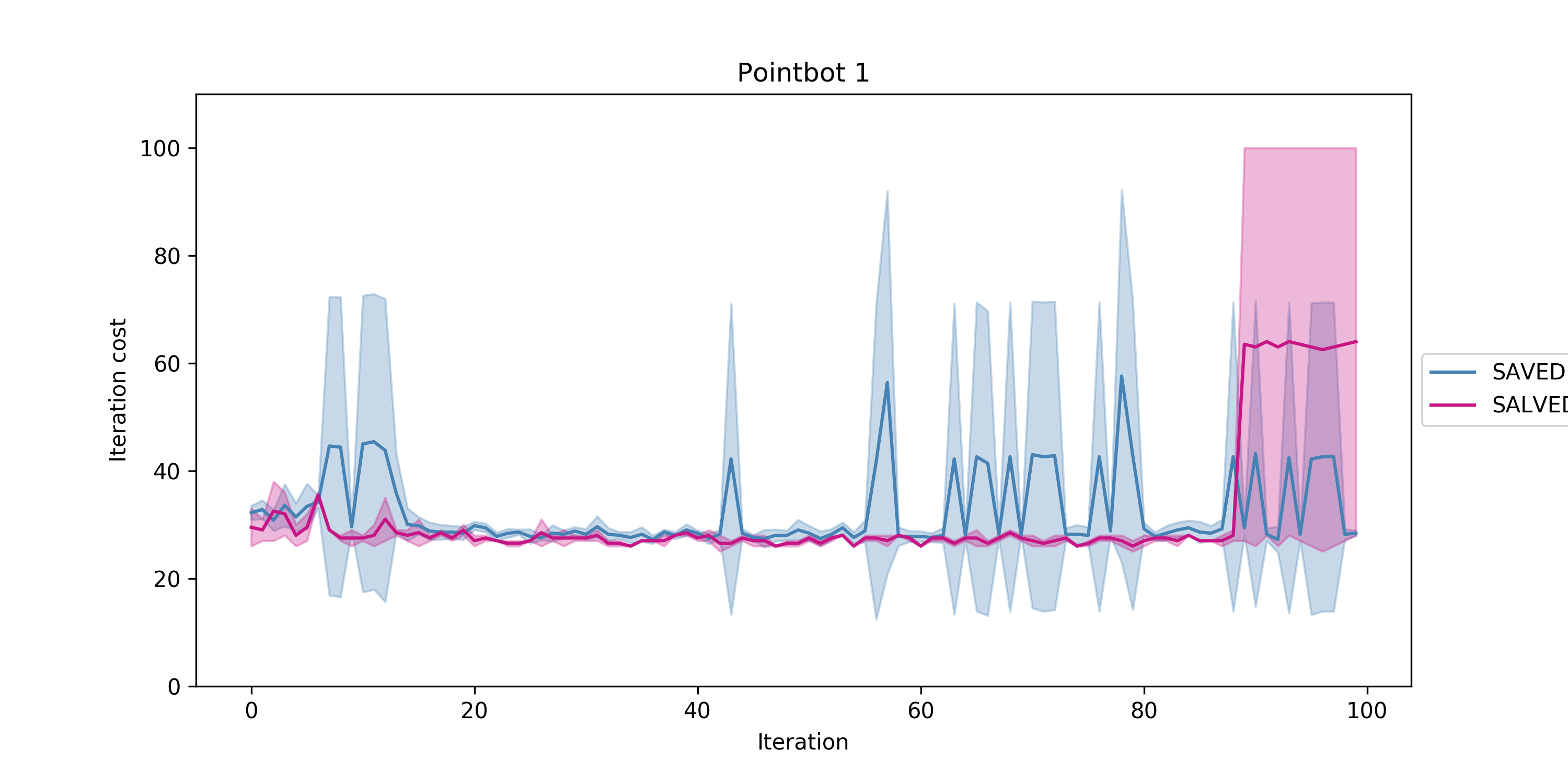}
  \caption{Pointbot 1}
  \label{fig:pb1-iter-cost}
  \end{subfigure}\hfil
  \begin{subfigure}{0.5\textwidth}
  \includegraphics[width=\linewidth]{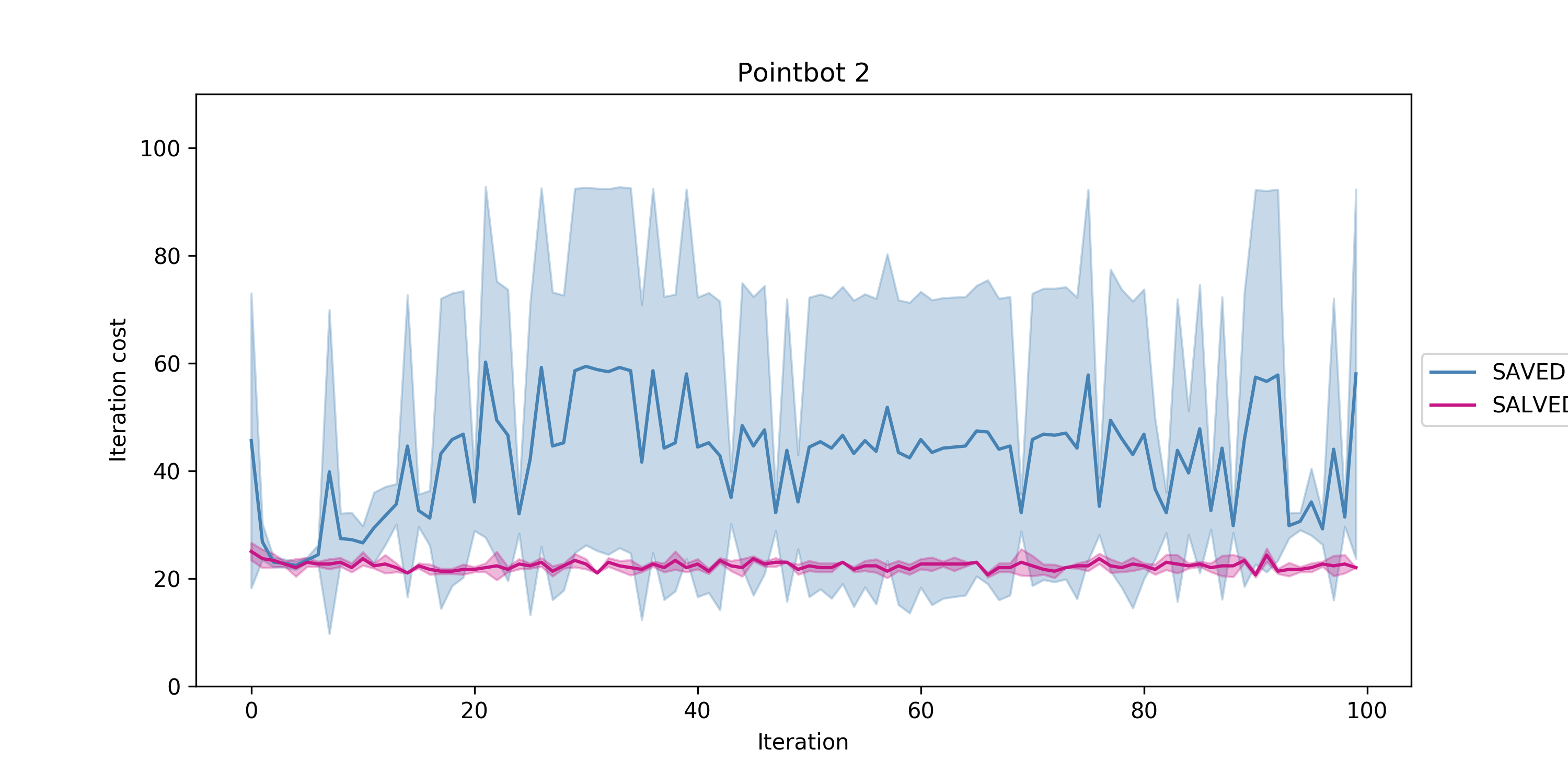}
  \caption{Pointbot 2}
  \label{fig:pb2-iter-cost}
  \end{subfigure}
  \centering
  \begin{subfigure}{0.5\textwidth}
  \includegraphics[width=\linewidth]{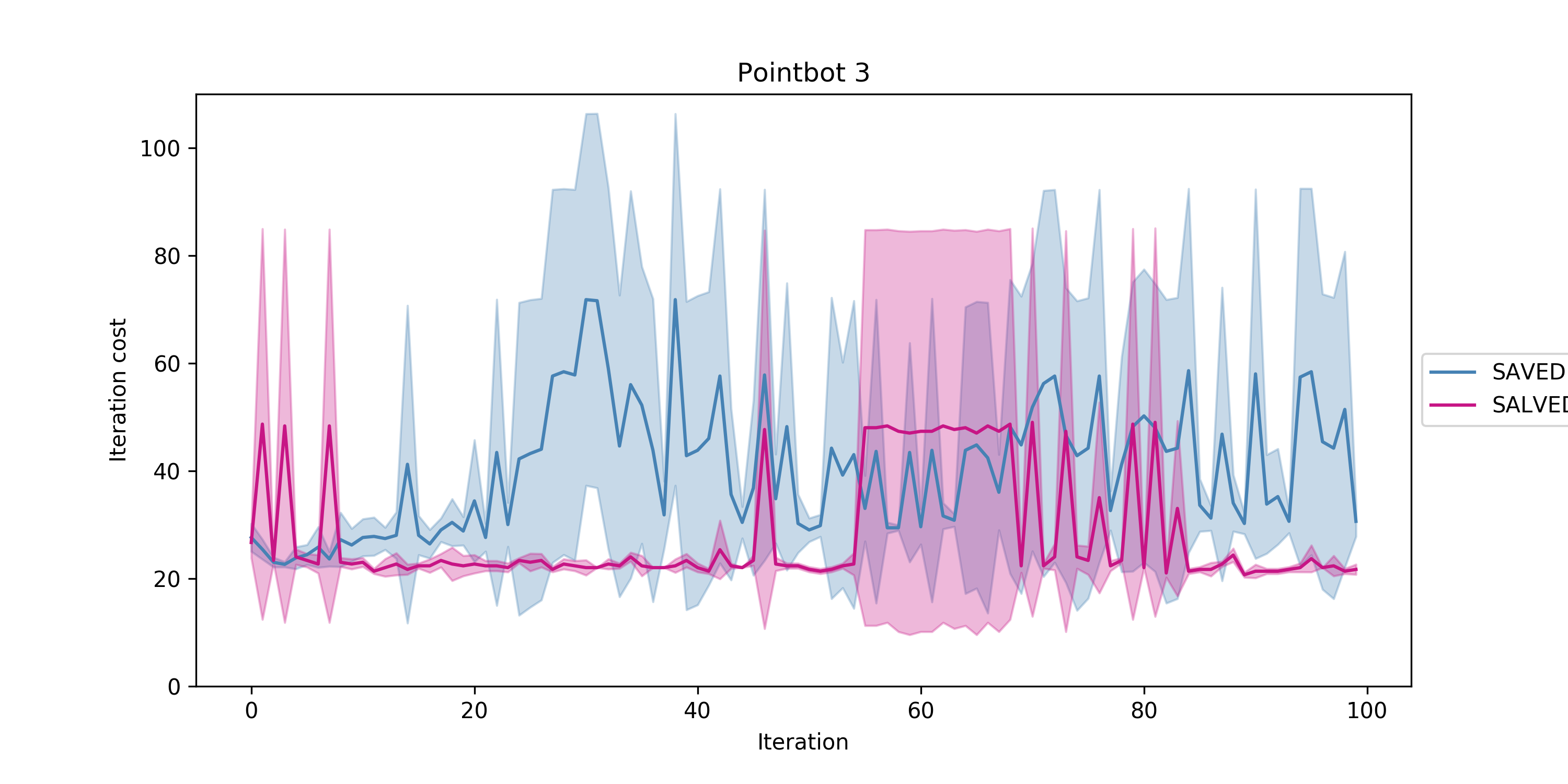}
  \caption{Pointbot 3}
  \label{fig:pb3-iter-cost}
  \end{subfigure}\hfil
  \begin{subfigure}{0.5\textwidth}
  \includegraphics[width=\linewidth]{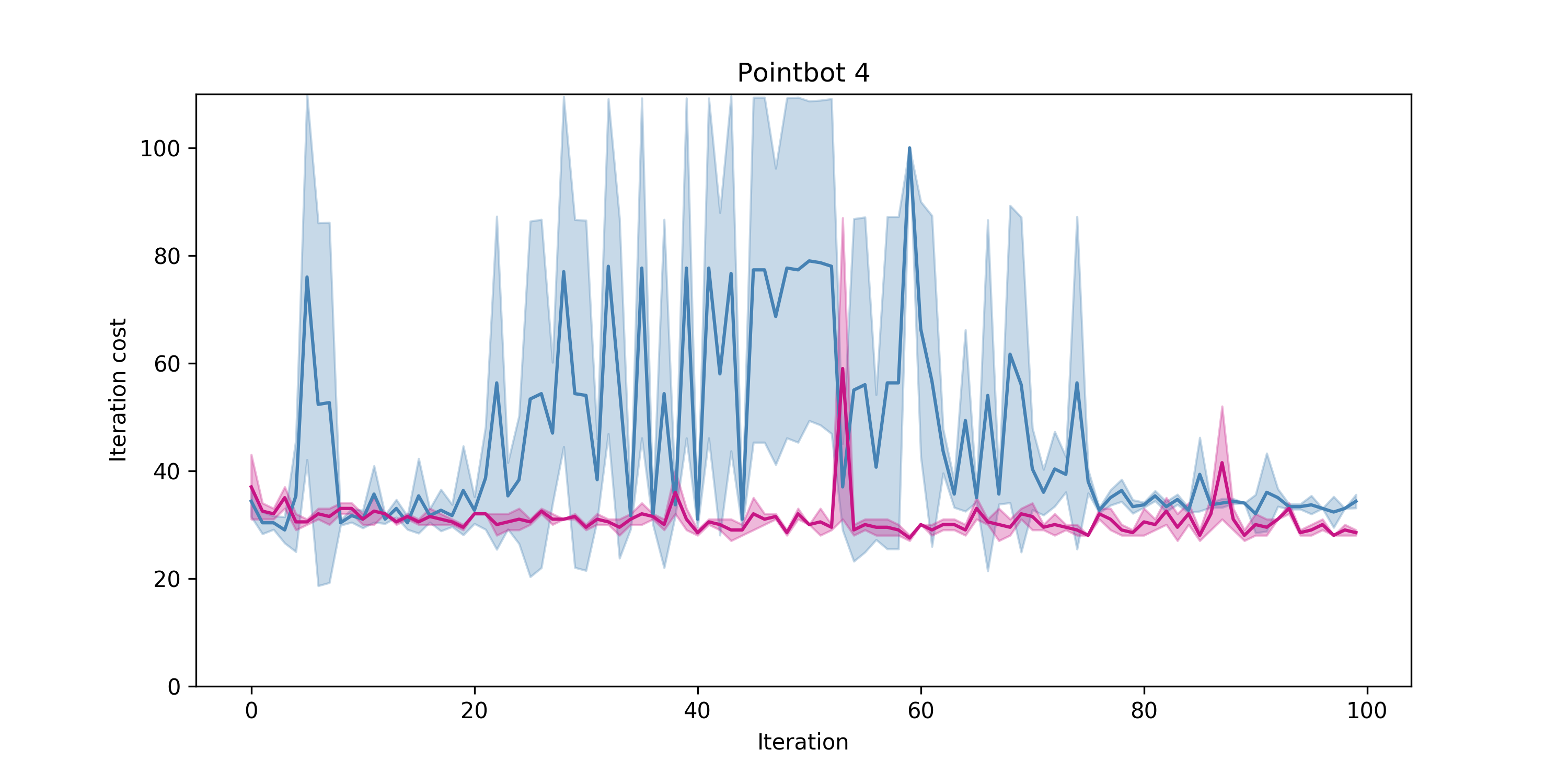}
  \caption{Pointbot 4}
  \label{fig:pb4-iter-cost}
  \end{subfigure}
  \caption{\textbf{Pointbot Navigation Tasks}: SALVED is evaluated on four Pointbot navigation tasks. Pointbot 1 has no obstacles or safety constraints, Pointbot 2-4 contain obstacles with increasingly more complex constraints. We observe that SALVED has a lower iteration cost than baselines throughout all four tasks and shows considerably more stability during training and evaluation.}
  \label{fig:iteration-cost}
\end{figure}

\begin{figure}[H]
  \centering
  \begin{subfigure}{0.5\textwidth}
  \includegraphics[width=\linewidth]{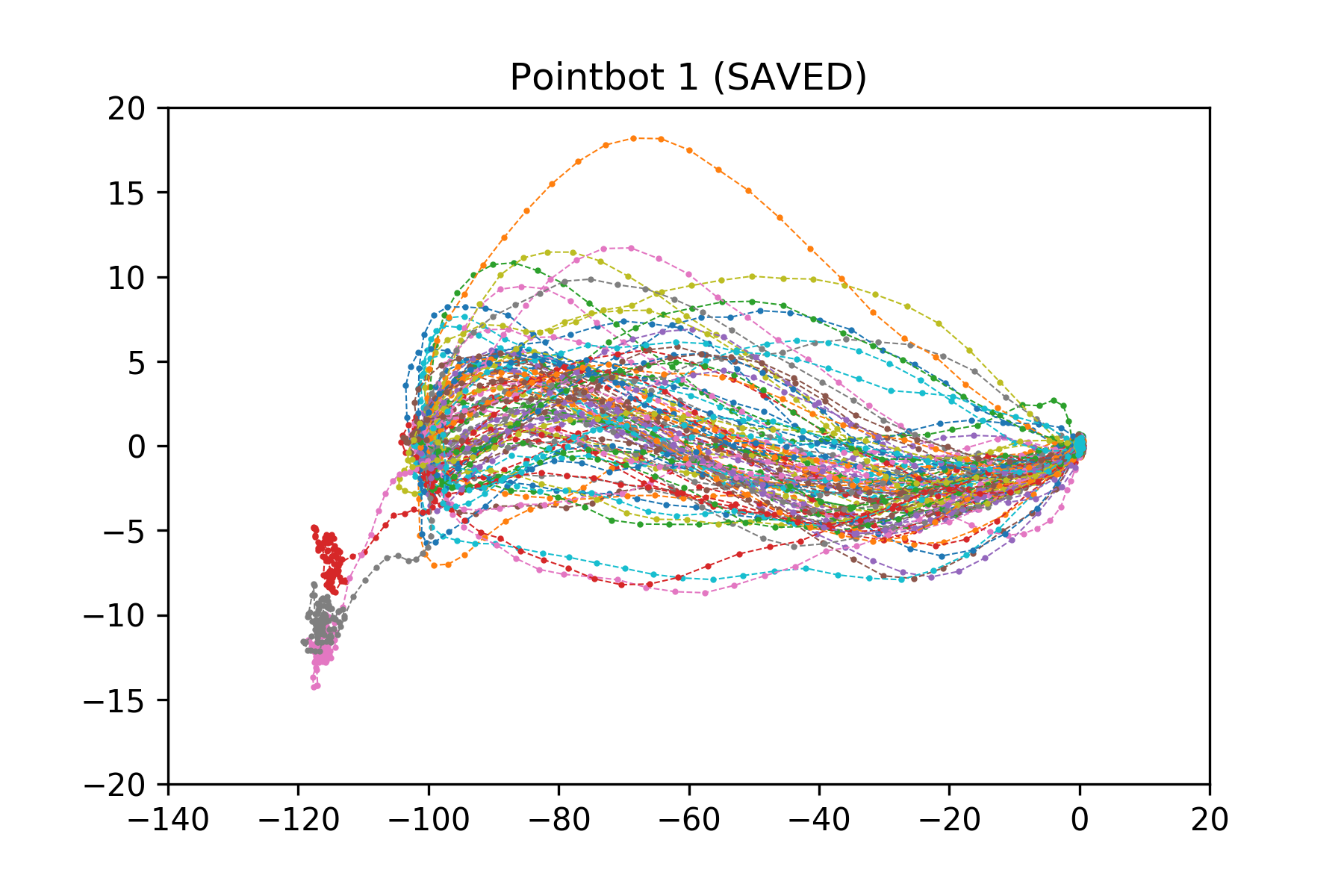}
  \caption{Pointbot 1 (SAVED)}
  \label{fig:pb1-saved-traj}
  \end{subfigure}\hfil
  \begin{subfigure}{0.5\textwidth}
  \includegraphics[width=\linewidth]{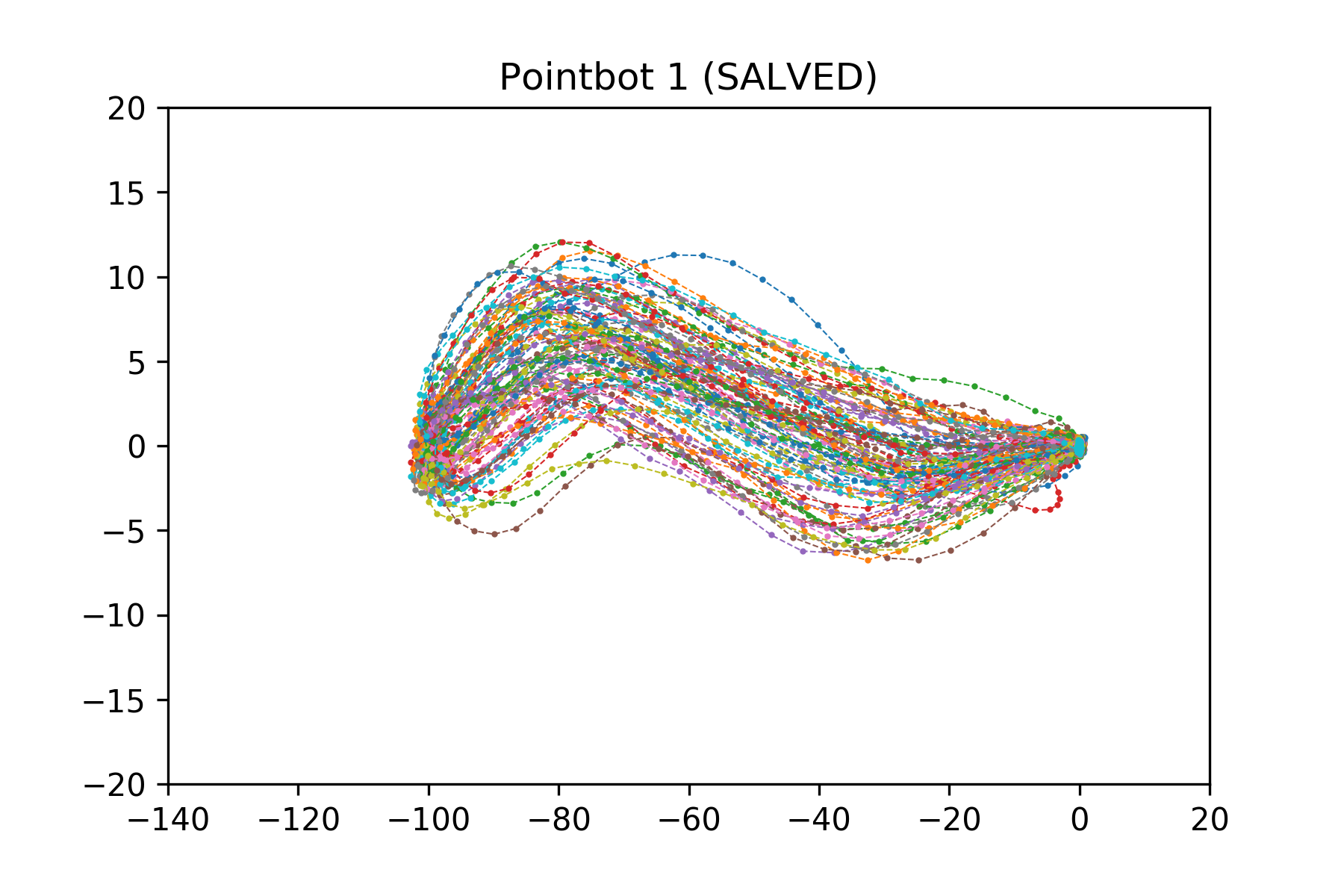}
  \caption{Pointbot 1 (SALVED)}
  \label{fig:pb1-salved-traj}
  \end{subfigure}

  \begin{subfigure}{0.5\textwidth}
  \includegraphics[width=\linewidth]{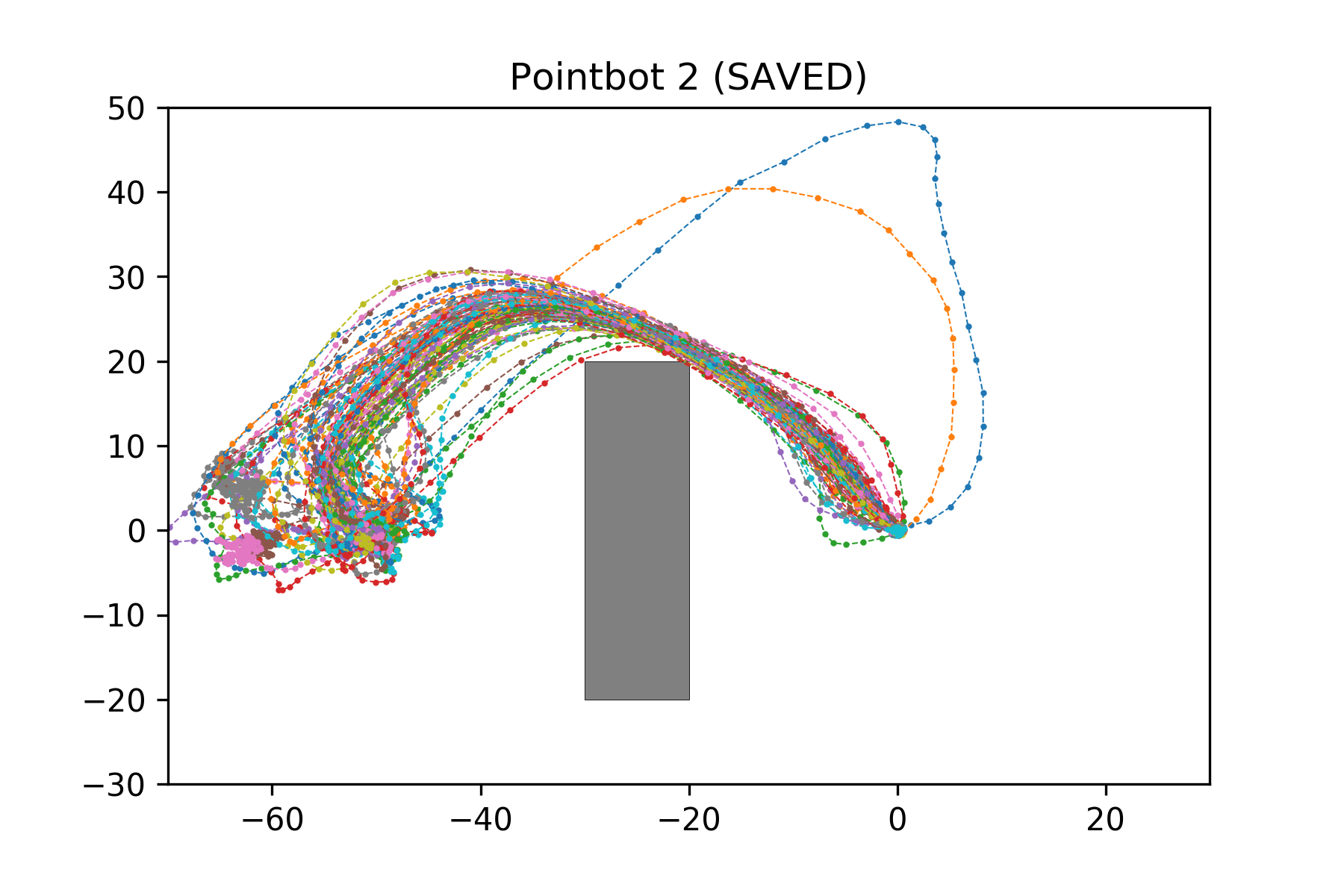}
  \caption{Pointbot 2 (SAVED)}
  \label{fig:pb2-saved-traj}
  \end{subfigure}\hfil
  \begin{subfigure}{0.5\textwidth}
  \includegraphics[width=\linewidth]{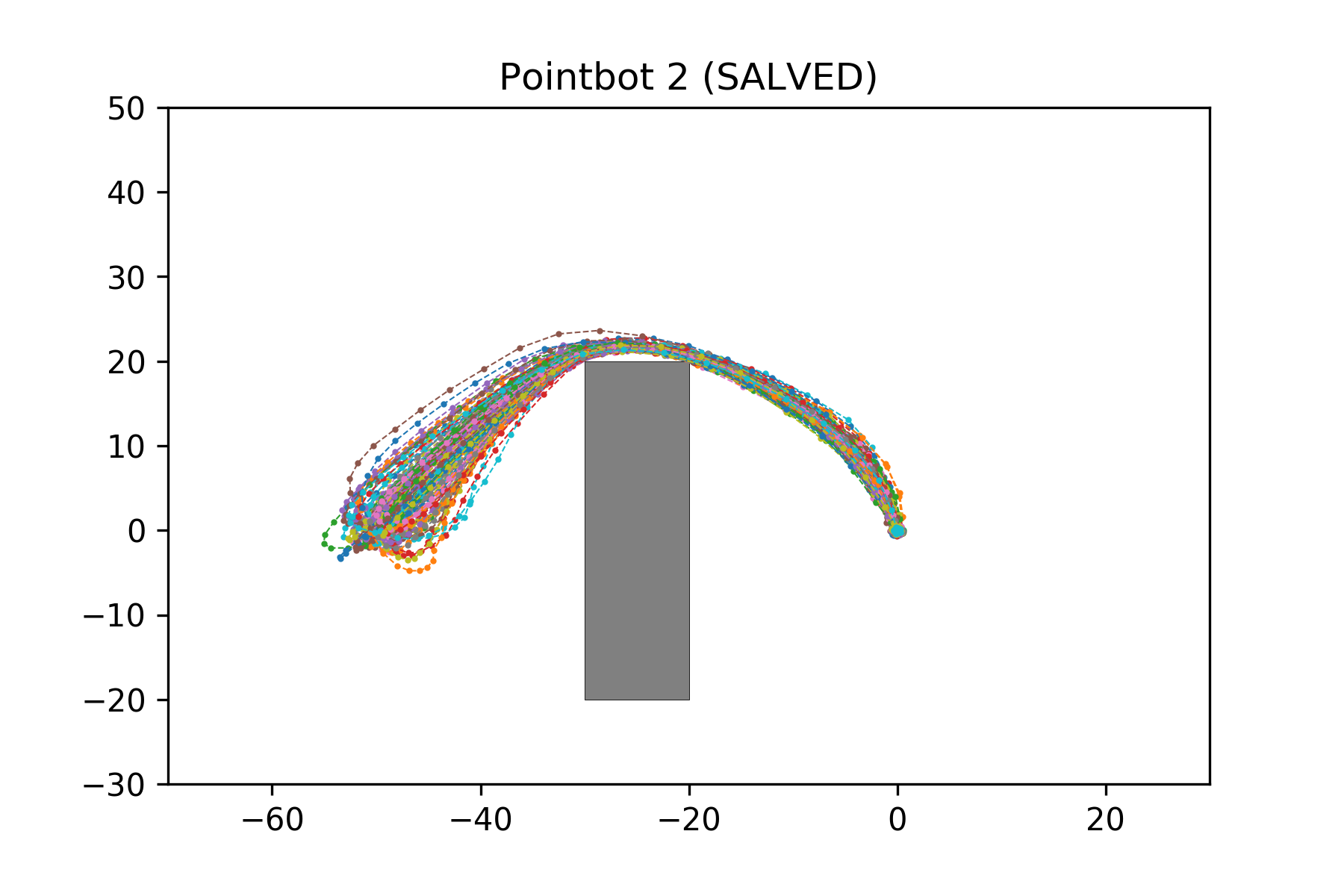}
  \caption{Pointbot 2 (SALVED)}
  \label{fig:pb2-salved-traj}
  \end{subfigure}
\end{figure}

\begin{figure}\ContinuedFloat
  \centering
  \begin{subfigure}{0.5\textwidth}
  \includegraphics[width=\linewidth]{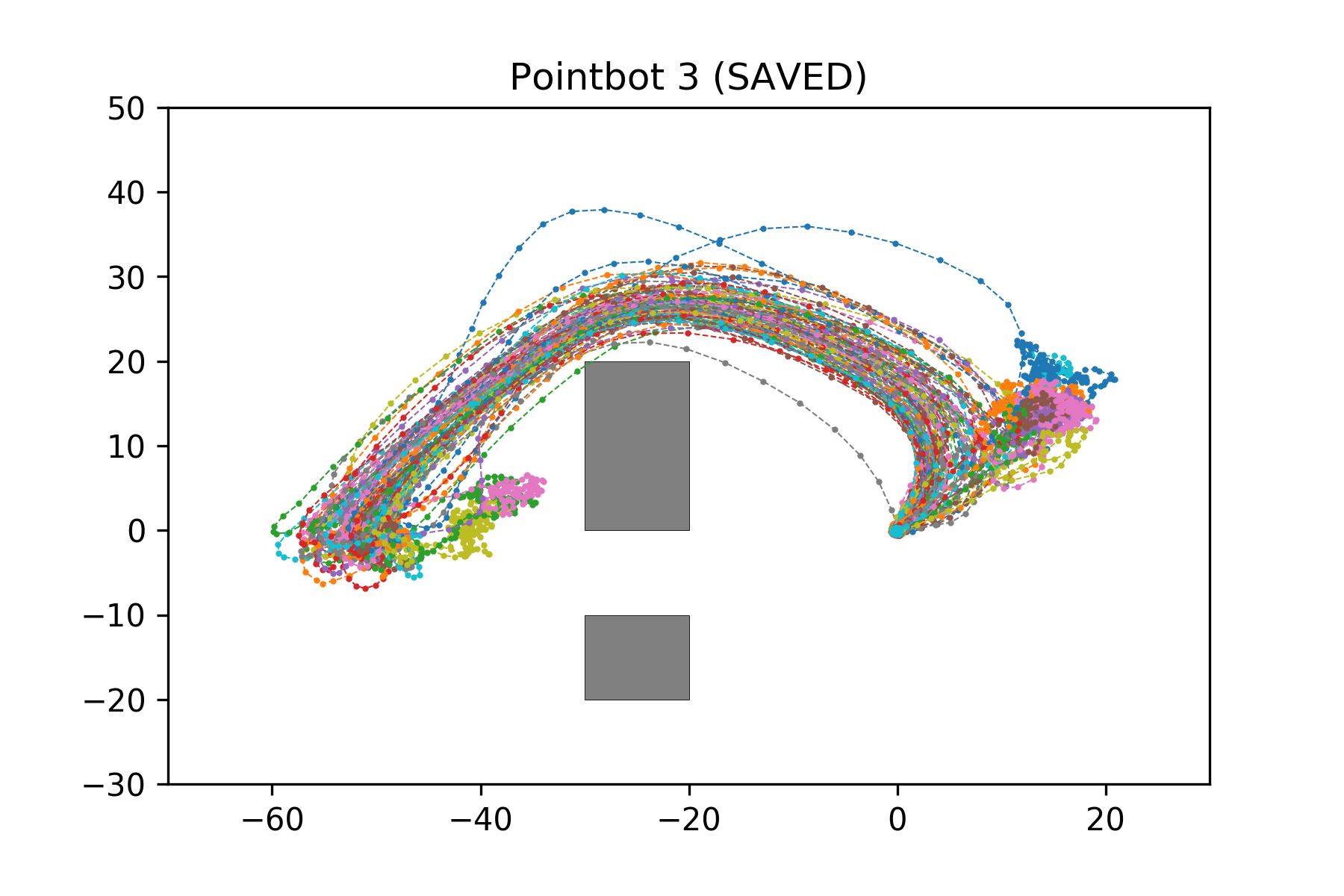}
  \caption{Pointbot 3 (SAVED)}
  \label{fig:pb3-saved-traj}
  \end{subfigure}\hfil
  \begin{subfigure}{0.5\textwidth}
  \includegraphics[width=\linewidth]{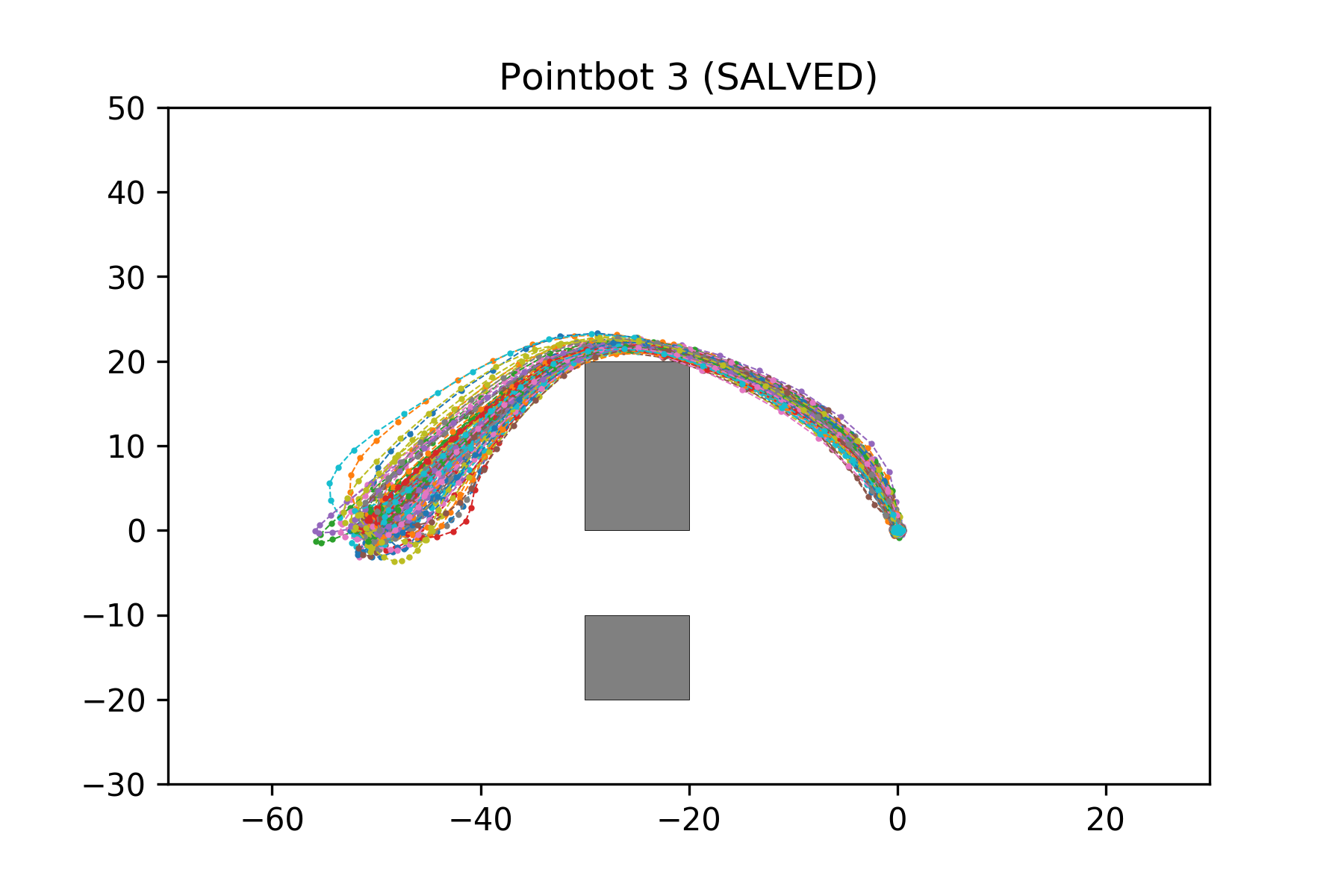}
  \caption{Pointbot 3 (SALVED)}
  \label{fig:pb3-salved-traj}
  \end{subfigure}
  \begin{subfigure}{0.5\textwidth}
  \includegraphics[width=\linewidth]{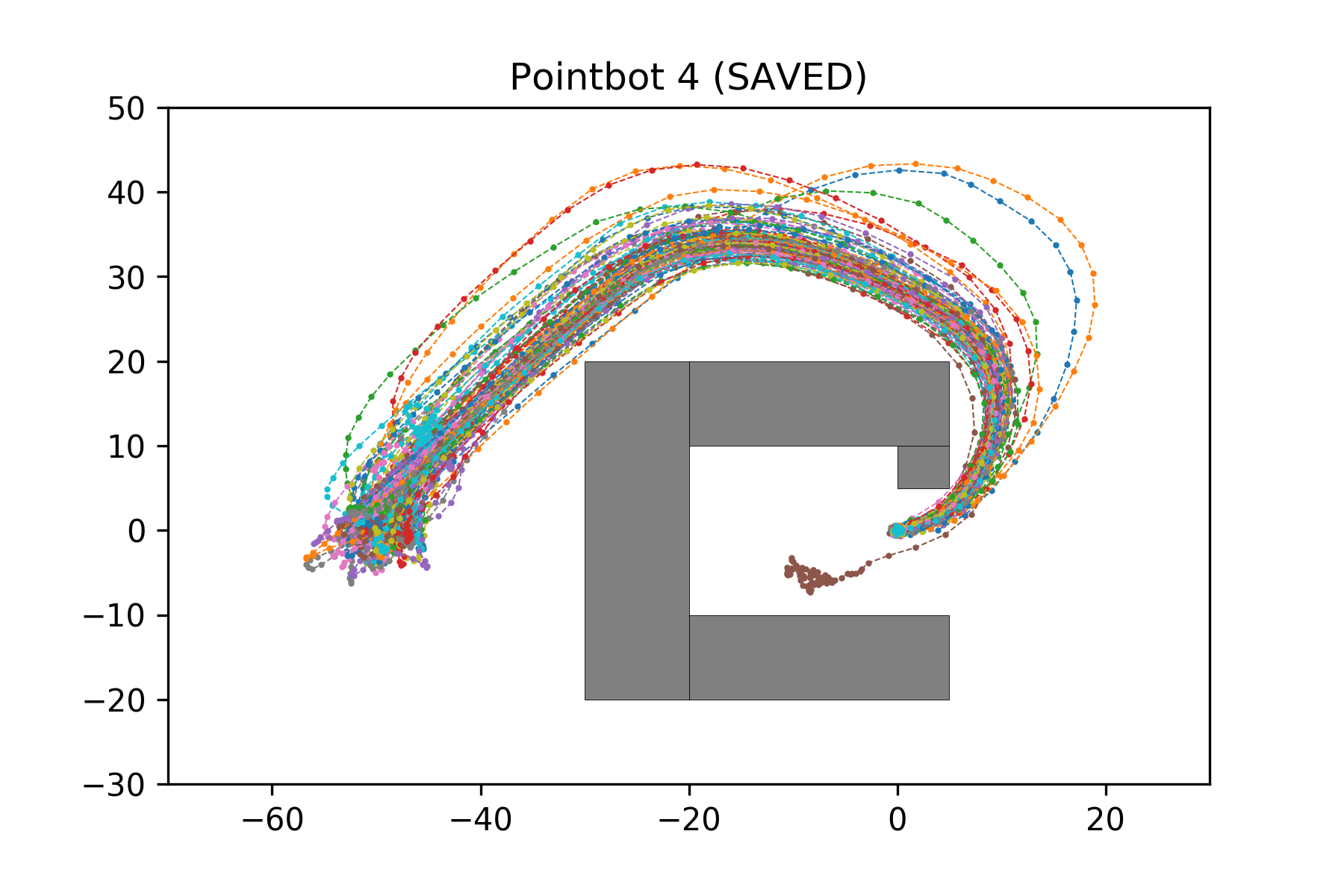}
  \caption{Pointbot 4 (SAVED)}
  \label{fig:pb4-saved-traj}
  \end{subfigure}\hfil
  \begin{subfigure}{0.5\textwidth}
  \includegraphics[width=\linewidth]{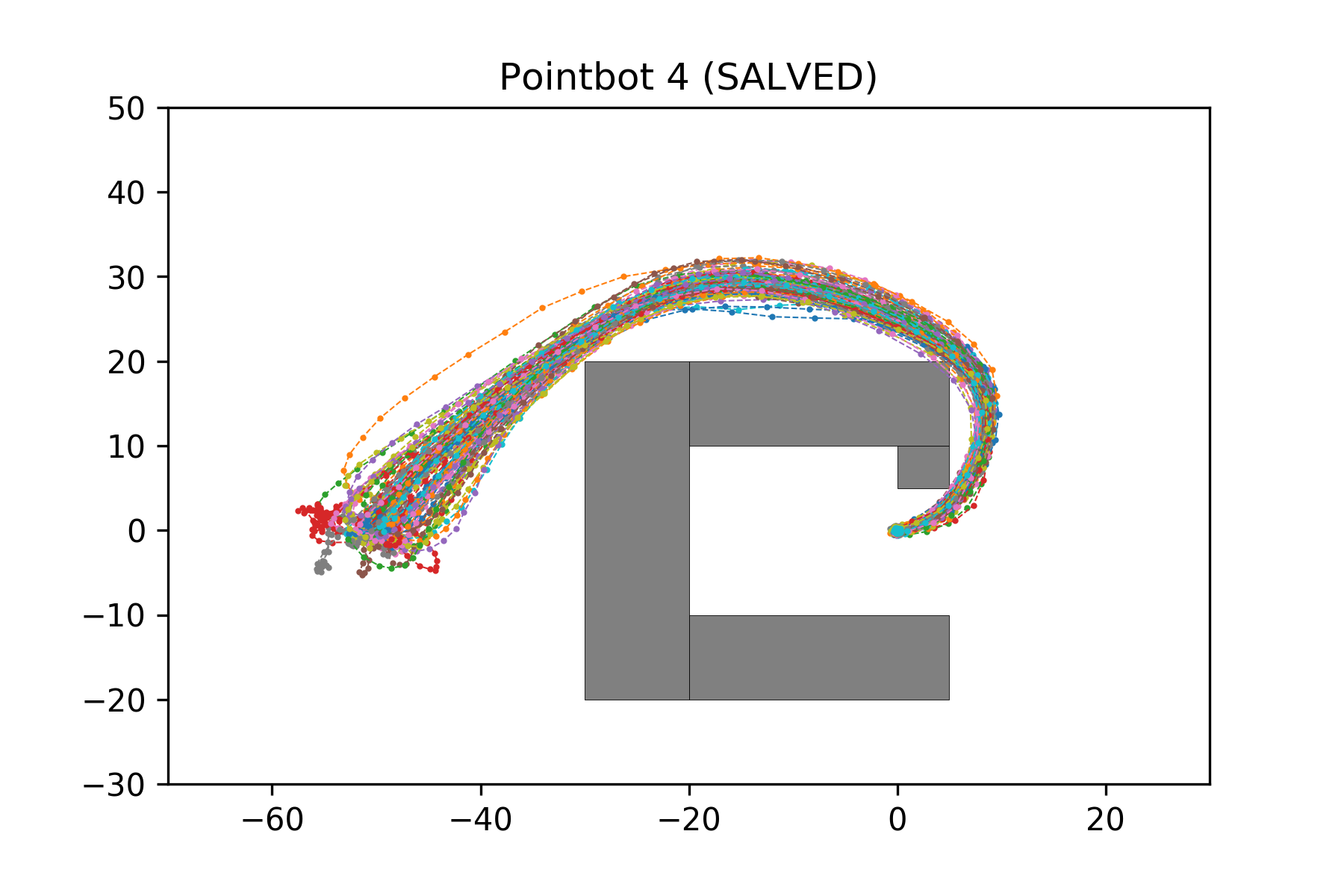}
  \caption{Pointbot 4 (SALVED)}
  \label{fig:pb4-salved-traj}
  \end{subfigure}
  \caption{The plots show evaluation trajectories combined across 100 iterations. The left plots show that the control policies learned by SAVED are considerably less optimal and have lower task completion rates across the experiments. The right plots demonstrate higher task completion rates and suggest that the policies learned by SALVED are significantly more optimal and consistent as the resulting trajectories are also more dense.}
  \label{fig:pointbot-eval-traj}
\end{figure}

\section{Conclusion}

In this paper, we present SALVED, a model-based reinforcement learning algorithm with provable stability guarantees by learning Lyapunov neural network value functions. We empirically evaluated SALVED on four simulated benchmarks and found that SALVED provided significant performance improvements over the baseline in all four environments. SALVED is able to maintain the sample efficiency of model-based RL algorithms such as SAVED and the iterative improvement guarantees of the Learning MPC framework, while providing additional Lyapunov stability guarantees. Results not only show higher task completion and constraint satisfaction rates, but also better-performing learned control policies with more optimal trajectories with less variance across runs and controllers that are less prone to local minima and catastrophic failures from constraint violations due to the stability guarantees from the Lyapunov-constrained value function approximation.

\bibliographystyle{IEEEtranN}
\bibliography{reference}

\end{document}